# Optical Switching Data Center Networks: Understanding Techniques and Challenges

**Xuwei Xue[1], Shaojuan Zhang[2]\*, Bingli Guo[1], Wei Ji[3], Rui Yin[3], Bin Chen[4], Shanguo Huang[1]**

[1]State Key Laboratory of Information Photonics and Optical Communications (IPOC), Beijing University of Posts and Telecommunications, 10 Xitucheng Rd, Bei Tai Ping Zhuang, Haidian Qu, Beijing, 100876, China
[2]IPI-ECO Research Institute, Eindhoven University of Technology, 5600MB Eindhoven, The Netherlands
[3]School of Information Science and Engineering (ISE), Shandong University, no. 72 Beihai road, Jimo district, Qingdao, 266237, China
[4]School of Computer Science and Information Engineering, Hefei University of Technology, 193 Tunxi Rd, Baohe Qu, Hefei, 230601, China.

E-mail: s.zhang4@tue.nl

**Abstract**

Relying on the flexible-access interconnects to the scalable storage and compute resources, data centers deliver critical communications connectivity among numerous servers to support the housed applications and services. To provide the high-speeds and long-distance communications, the data centers have turned to fiber interconnections. With the stringently increased traffic volume, the data centers are then expected to further deploy the optical switches into the systems infrastructure to implement the full optical switching. This paper first summarizes the topologies and traffic characteristics in data centers and analyzes the reasons and importance of moving to optical switching. Recent techniques related to the optical switching, and main challenges limiting the practical deployments of optical switches in data centers are also summarized and reported.

## Introduction

Dater centers (DCs), consisting of tens thousands of servers connected by large switching networks, provide the infrastructure for online applications and services such as cloud computing, social networks, file storage, and web search [1]. The topology of data center networks (DCNs) plays significant roles in determining the communication bandwidth between servers, the flow completion time and fault tolerance [2]. The design of the DCN topology is thus to build a robust network that provides the high bandwidth links and low (typically hundreds of microseconds) flow completion time across servers with low building cost and power consumption. Due to the various hosted application and services, the DC traffic, consisting of the tenant-generated interactive traffic, deterministic traffic and traffic with deadlines, is a mix of several classes with differentiate characteristics [3]. Thus, the DCN infrastructure should be effectively and efficiently utilized to provide high performance and quality access to the variety of services and applications deployed in DCs.

With the escalation of traffic-increasing applications, such as high-definition streaming, Internet of Things and cloud computing, traffic growth in DCs exceeds the bandwidth growth rate of application-specific integrated circuits (ASICs) based electrical switch [4]. The ASIC switches are expected to hit the bandwidth bottleneck in two generations from now, because the Ball Grid Array (BGA) packaging technique is hard to increase the pin density [5]. Moreover, the hierarchical network topologies based on electrical switches further deteriorates the bandwidth provisioning, especially at the top-layer switch, stringent increasing the flow completion time. As future-proof solutions supplying high bandwidth, switching traffic in the optical domain has been considerably investigated to overcome the bandwidth bottleneck of electrical switches. The optical switches with high bandwidth, because of the optical transparency, are independent of the data-format and data-rate of the traffic [6]. Moreover, switching the traffic in the optical domain eliminates the power-consuming optical-electrical-optical (O-E-O) conversions. Migration of the traffic switching from electrical domain to the optical domain also removes the electronics circuits for dedicated various-format modulation at transceivers, thereby, significantly reducing cost expenses and data processing delay [7]. To date, the optics and networking communities have proposed many solutions on optical switches with milliseconds to nanoseconds switching configuration time, and variety switches based DCN topologies.

To practically deploy optical switches in DCNs, there are still several challenges that need to be addressed. First, to fully utilize the nanoseconds-level hardware switching time, a corresponding switch control mechanism is required to manage the optical switches in nanoseconds time scale to fast switch the traffic data [4]. Second, the conflicted packets at optical switches would be dropped, as no optical buffer existing to store the conflicted packets. This would result in high packet loss when packet contention happens at the optical switch nodes. Thus, packet contention resolution is another unsolved challenge to practical deploy the optical switches in DCNs [8]. Third, in optically switched network, new optical connections are generated every time as the switch reconfiguring. This requires that the receivers having to continuously adjust the local clock to properly sample the incoming packets and then recover the data [9]. The longer this recovery process takes, the lower the network throughput will be, particularly for the intra data center scenarios where many applications produce short traffic packets. Four, the multi-tenant services and applications with various data flows impose their own set of heterogeneous traffic requirements to the DC infrastructure [10]. Thus, a reconfigurable and highly flexible connectivity for DCN is required to provide the customized network frameworks to the various applications.

In this paper, we present a review of optical switching techniques capable of meeting the requirements of the next generation of large-scale data center networks. We start with a summarization of current data center traffic characteristics and topologies that revels the requirements of improving network performance in terms of both switch nodes and network topologies. To overcome the bandwidth limitation and multi-tier architecture of electrically switched networks, optical switching techniques have been proposed and investigated to replace the current electrical switches. We then review the technologies involved in the optical switch fabrics and the switch based optical topologies. The challenges of limiting the practical deployment of optical switching data centers have also been proposed to inspire researchers to propose more solutions. Finally, we summarize our conclusions.

**Data Center**

A data center is a physical facility, dedicated space within a group of buildings that operators use to compute, store and forward large amounts of traffic data. A data center's design is based on a computing network and associated components, such as storage and telecommunications that enable the operations of housed applications and services [1]. DCs consist of servers, switches, storage systems, routers, firewalls and application delivery controllers. Because these components are operated to process and store the business-critical data, stability is critical in the design and build of data centers. Together, they provide [11]: 1) Computing resources. The servers provide the computing, local memory, storage and network connectivity to drive the services and applications that are the engines of the network; 2) Storing infrastructure. Applications and services' data is the fuel of data centers. Storing systems are utilized to reliably hold this valuable information; 3) Network infrastructure. This system connects internal servers, storages, and external connectivity to end-users. Based on these infrastructures, data centers of worldwide enterprise IT are built to operate business applications and services that include but not only limited to [12]: 1) Productivity applications; 2) Email and file-sharing; 3) Database and analytics; 4) enterprise resource planning (ERP) and Customer relationship management (CRM); 5)

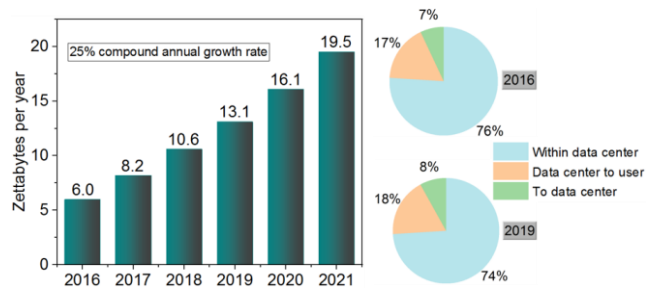

Figure. 1. Data center traffic growth and distributions [15].

Communications and collaboration services; 6) Artificial intelligence, big data and machine learning.

In the 5G era, data centers are evolving to provide high-speed and quality access, triggered by the new emerging cloud computing, 5G smart services and Internet of Things (IoT) applications. In this era, the digital ecosystem chain covers centralized data center network, edge computing, and terminal-device and everything in between [13]. This requires the DC infrastructure migrating from conventional on-premises physical servers to virtualized infrastructure that supports workloads across pools of physical infrastructure and into a multi-cloud environment[14]. Moreover, the escalation of traffic-boosting applications and the scale out of powerful servers have heavily increased the traffic volume in DCs. As shown in Fig. 1, by the end of the year 2021, annul global data center IP traffic is projected to reach 19.5 Zettabytes, which represents almost four-fold increase from the year 2016 [15]. About three-quarters of the business and consumer traffic flowing in data centers resides within the data center network. As the core hub of the digital ecosystem, data centers are phenomenally growing in complexity and size to evolve infrastructure with high network performance and to satisfy the keep-increasing traffic, playing the pivotal role in this innovative era and cross-generation evolutionary.

*1.1 Data Center Network Topology*

Data center networks (DCNs) establish and connect the entire network-based equipment and devices within the data center facility to enable a reliable interconnection. The DCNs thus ensure that the inside facility nodes can communicate and transfer data between each other and to the external users [16]. Existing electrical switched based DCN architectures are classified into two categories: server-centric and switch-centric architectures [17]. In server-centric networks, switch nodes are utilized as cross-connects, and routing intelligence should be placed on servers, where multiple Network Interface Card (NIC) ports are used per server. In switch-centric networks, routing intelligence is placed on switch nodes and each server usually uses only one NIC port to connect to the network.

The major advantage of switch-centric networks with the separation of communication and computation is that they are based on proven traffic forwarding and routing technologies available in commodity switches (e.g., Ethernet switches), such as IP broadcasting, link-state routing, and equal-cost multi-path forwarding [18]. Although a number of architectures in server-centric design have been proposed exploiting low-cost switches, the switch-centric based architectures are the mainstream scheme for the DCNs [19]. For instance, the multi-tier tree-like architectures continues to be the most widely deployed, and the fat-tree and leaf-spine are the most promising architectures in terms of robustness, scalability and cost. All these architectures are switch-centric design.

The multi-tier design of DCNs comprises of hierarchy of switches layers as depicted in Fig. 2. The leaves of the network tree form the access switching layer. Switches in this layer are usually top of rack (ToR) switches at low-cost, connecting servers (typically 20-40 servers) locating in the same rack. The middle tier of the network tree forms the aggregation switching layer, and the root of the network tree forms the core switching layer. In the multi-tier network, multiple racks are grouped together into one cluster. The intra-cluster and inter-cluster communication are handled by layers of aggregation switches and core switches, respectively. As illustrated in Fig. 2, these access switches are connected through optical links to the aggregation switches to forward intra-cluster traffic. The inter-cluster traffic data is forwarded by the aggregation switches connected to the core switches. One or more border core layer switches provide connectivity between network infrastructure and users.

This multi-tier tree-like architecture features high fault tolerance enabled by the extensive path diversity even under failures. The main drawback of this tree-like design is the less than 1:1 oversubscription because of prohibitive costs. The 1:1 oversubscription means that any server can communicate at full bandwidth of their network interface with other arbitrary servers [20]. Due to the linear increasing costs associated with the scaling of link bandwidth and the port density of conventional electrical switches, building a tree-like scheme with 1:1 oversubscription would be prohibitively expensive for large-scale DCNs [21]. Therefore, practical oversubscription in this topology typically ranges from 8:1 to 3:1. Under the high oversubscription, if more traffic is generated at the certain time on the active link, the large-stocked traffic exceeding the routing table entries will significantly increase the transmission latency [22].

As schematically illustrated in Fig. 3, the fat-tree and leaf-spine data center network topologies have been evolved as the typical DCN architectures. As compared to multi-tier tree-like networks, the fat-tree architecture supports the use of commodity, identical switches in all switching layers, thereby decreasing cost in multiple-times. As depicted in Fig. 3(a), servers in the same rack are connected to a ToR switch, and this ToR switch is connected to a set of aggregation switches. At the root of fat-tree topology, a set of core switches are connected to aggregation switches in each cluster. In fat-tree network, aggregation switches and ToR switches of every cluster provide sufficient bandwidth for intra-cluster traffic forwarding, such that servers in the same cluster can communicate with each other at full speed [23].

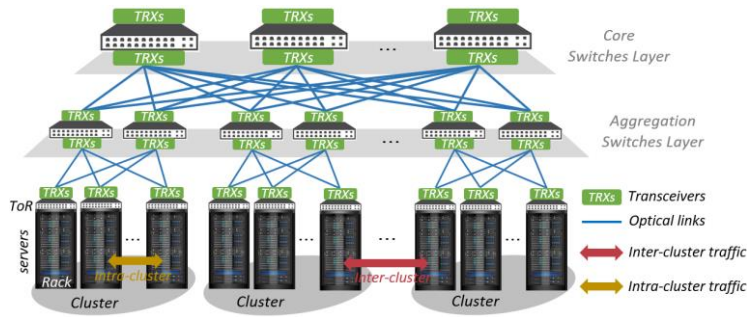

Figure. 2. Multi-tier tree-like data center network.

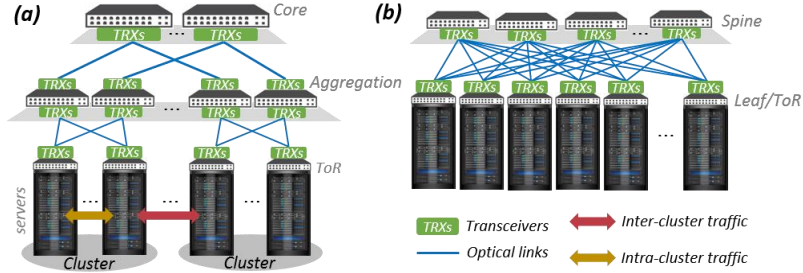

Figure. 3. (a) Fat-tree and (b) leaf-spine DCN topology.

As a typical DCN topology, especial for the big data industry, the leaf-spine design has only two layers as shown in Fig. 3(b), the spine switching layer and leaf switching layer. The spine layer is the backbone of the network, where every leaf switch is interconnected with each-and-every spine switches. The leaf layer consists of ToR switches that connect to rack servers. With such a leaf-spine network topology, no matter which server is connected to which leaf switch, its traffic always crosses the same number of devices to arrive at the other server. Because a packet only needs to hop to a spine switch and another leaf switch to reach its destination, this guarantees the packet completion time at a predictable level [24]. Another benefit of leaf-spine topology is the capability of adding additional hardware and bandwidth. When oversubscription occurs on a certain link, an additional spine switch can be added and bandwidth can be thus extended to every leaf switch to reduce the oversubscription [25].

## 1.2 Data Center Traffic

Variety of applications and services are deployed on data centers benefiting from the flexible and cost-effective access to scalable storage and compute resources [12]. Data center traffic is often a mix of several classes with different service priorities and application requirements, which intimately determines the topology, scale, and even the technology selection of the DCNs [26]. To this end, a full understanding of the data center traffic characteristics is extremely necessary before putting efforts into the topology identifying and technology selection. Traffic characteristics, such as flow types, size and arrivals are highly correlated with applications. It is thus relatively hard to conclude a general pattern about traffic characteristics, given such strong dependency on data center applications. Some interesting findings include the following.

*Mix of Flow Types:* Data center traffic is a mix of various flow types and sizes [27]. A traffic flow here is specified as an established link between any two servers. User interactions like soft real-time applications such as Web search can create interactive traffic which are latency-sensitive flows that are usually short and should be forwarded with high priority [28]. Examples include short messages (100 KB to 1 MB) and queries (2 to 20 KB) [29]. Throughput-oriented flows require consistent bandwidth, but they are not sensitive to delay. These flows length ranges from moderate transfers (1 MB to 100 MB), such as one created by data computing applications (e.g., MapReduce), to long-running background flows, such as delivering large volumes of data across network sites for data storage [30]. Deadline flow has to complete the flow prior to the certain deadlines [31]. The size of deadline flow is either pre-known or a good estimate can be typically drawn [32]. The flow deadline could be either hard or soft which implies how value of flow completion decrease as time passed [33]. The hard deadline means zero value once the deadline has passed, while the soft deadline implies that it is still profitable to complete the flow and the value drops as the time passes away from the software deadline.

*Traffic Burstiness:* Burst is a common characteristic for data center traffic. Hardware offloading features, such as Interrupt Moderation and Large Send Offload that release the traffic workload of CPU, can lead to high burstiness [34]. Transport control in software, such as TCP slow start, can also create burst traffic when sends together a large window of flows [35]. In a burst environment, packet loss due to higher burstiness has been found more frequent at the network edge nodes like ToRs [36]. Burstiness can also lead to higher average buffer queue occupancy and thereby increasing packet drops probability and flow completion time due to buffers overflowed [37, 38]. Moreover, high-burst traffic can deteriorate the buffer utilization in the memory shared switches when a certain switch port exhausts the shared buffer resulting from the long received bursty traffic [39].

*Unpredictable Traffic Matrix:* The various applications and services running on a data center create variety flows with different properties. Most bytes are delivered by large flows and most flows are short (less than 1 MB). The data center monitored in [40] shows that 99% of flows are shorted than 100MB and more than 90% of bytes are forwarded in flows between 100MB and 1GB. The flow arrival rates and distributions are also determined by the data center applications. The inter-arrival time of median flows in Facebook's DCs are between 2 ms to 10 ms for a single server (100 to 500 flows per second) while Microsoft finds the median arrival rate for the intra-cluster traffic to be 100 flows per millisecond [28, 41]. Reference [42] reports between 100 to 10000 flow arriving at the switch per seconds in different private and educational data centers. Such variety of flow length and flow arrival rate can create a fluctuating and unpredictable traffic matrix which makes it hard to operate capacity planning and perform traffic engineering in a long-term scale to improve the network performance.

Data center infrastructures can be shared by multi-tenants and hosted applications offering various services to the network end-users. Traffic analyzation is then a necessary task in DCNs to guide the efficient use of the network resources. Knowledge of various traffic requirements and characteristics, as can be seen from the above empirical results, can help us design transport protocols and even network topologies to more efficiently use network resources.

*1.3 Design Requirements for Data Center Networks*

So much more than just a warehouse for servers, the data center is a sophisticated data networking environment that offers reliable and high-quality services to its users and customers, empowering them to design new infrastructures to enable their business forward. When design a new data center infrastructure, it's important to comprehensively consider the following summarized requirements that have a direct impact on performance.

*Capacity:* Recently, the escalation of traffic-boosting applications and the scale-out of powerful servers have significantly increased the traffic volume within the data centers. As reported in [15], the annual global data center traffic has reached 10.4 Zettabytes by the end of the year 2019. Consequently, each aggregation switching node in the data center networks has to handle multiple to hundreds of Tb/s traffic. In addition, the traffic inside the modern data centers is expected to increase in the 5G era with a very high compound annual growth rate. Consequently, future data centers require ultra-high capacity networks to interconnect the infrastructure resources [43].

*Scalability:* Network scalability is the ability to easily subtract or add more storage and compute resources. For 'the old days' of on-premise DCNs, scalability was incredibly slow, costly, and difficult to manage. In 'the new days' of 5G era, the majority of network routing functions, physical network access, and innovative applications such as cloud computing and machine learning are typically deployed in centralized large-scale cloud data centers, while the Internet of Things device access and corresponding services platform, as well as diversified third-party applications are distributed and located at small-scale data centers [44]. Thus, the new scheme used to scale (add or subtract) the present built data centers to suit the new 5G fashion should be designed in a time-and-cost efficient manner.

*Flexibility:* Flexibility is increasingly important because many applications and services are dynamically deployed that require elastic network resource to efficiently accommodate them in the data center. The promising solution is to flexibly slice the network infrastructure following the virtualization strategy in a fully manageable and operable way to map the various applications' requirements. For the virtualized data center network, each infrastructure component (such as computing, storage and network) and its connections can be virtually recreated to slice the network [45]. Supervisory software creates these virtual components as various applications are required. In addition, virtualization enabled flexible data center often requires less power and space than a traditional data center. It can also be simpler to automate and update than the traditional data center [46].

*Intelligence:* The intelligent engines that are easier to use and configure with high efficiency, combined with network topologies and diverse domain knowledge of traffic characteristics, are required for modern data centers to quickly learn valuable information and execute targeted strategies from the massive amounts of data traffic generated by various applications. These intelligent engines should enable the data centers to provide rich platform services and application programming interfaces (APIs) with pre-integrated artificial intelligence (AI) services, search capabilities and graphics engine, as well as APIs in common fields such as voice, visual and language processing. These intelligent platforms and general pre-integrated services should work closely with the heterogeneous computing hardware such as field-programmable gate arrays (FPGAs) and graphics processing units (GPUs) to implement the application performance in-depth optimization [47].

*Effectiveness and Efficiency:* The traffic-boosting applications hosted in modern data centers impose stringent requirements in terms of packet loss, latency/jitter and throughput to the network infrastructure. For instance, the professional audio/video services and automatic driving applications require zero packet loss, low and bounded latency performance which is named deterministic quality of service (QoS) [48]. In addition, a data center represents a significant investment with the huge costs of hardware and software installation. Moreover, the cost of a DC's cooling and power typically is higher than the cost of IT devices inside it [49]. This is because components of modern data centers, including powerful servers, electrical switches and related network equipment, packages more transistors on the chip and more power-hungry chips in a smaller footprint. With the aim to satisfy these stringent network performance, cost/power-efficient data center networks featuring high bandwidth, providing extremely lower packet loss and latency performance should be investigated to host a broad range of mission-critical and latency-sensitive applications.

*1.4 Challenges for Data Center Networks*

Driven by the emerging of traffic-boosting applications and the scaling-out of powerful servers, more stringent requirements as abovementioned are imposed on the data centers with variety traffic characteristics. Current data center networks based on electrical switches are organized in a hierarchical topology, which is challenged by the bandwidth bottleneck and poor power efficiency to deliver the necessary and high quality of services [50].

*Electrical Switch:* The electrical switches double their bandwidth roughly for every two years at the same cost accordingly to Moore's law [51]. This allows data centers to keep up with the network bandwidth demands while maintaining the relatively steady and low network cost over the passing years [52]. However, the move towards traffic-boosting applications and powerful servers will greatly boost the demands of network bandwidth. Meeting the requirements of higher network bandwidth especial for the aggregation switch nodes, would greatly inflate costs.

Due to the limited number of high-speeds pins available on the switch chip and the limited number of connectors on the front panel of the rack unit, the bandwidth of electrical switches is expected to hit the bandwidth bottleneck soon [5]. Furthermore, the electrical switches consume the power proportional to the data rate, as the switch dissipates energy with every bit transition [53]. With the speed scaling-up requirements, the electrical switches based DCNs face the stringent pressures on the power-consumption. Despite new technologies based on multi-tier packaging, monolithic integration and Silicon Photonics (SiPh) are being investigated, several challenges, however, still have to be solved before these technologies become viable [54, 55]. For instance, the high manufacturing (including both packaging and testing) costs, not to mention the complexity of packaging a large number of fiber coupling and external laser sources. Even if these issues were able to be solved, these technologies will ultimately hard to keep increasing the transistor density limited by the CMOS scaling [56].

Hierarchical Network Topology: One of the performance deteriorations appeared with hierarchical data center topology is oversubscription that could dramatically disrupt the network performance [21]. Oversubscription is the ratio between the total uplink bandwidth to the servers' bandwidth at the ToR switch layer. Thus, due to the hierarchical network topology, as moving up to aggregation and core layers, the number of servers (and thus the bandwidth) sharing the uplink bandwidth increases and, hence, increases the oversubscription ratio, resulting in bandwidth bottlenecks at aggregation/core layers. Oversubscription limits the server to server capacity, especial for servers locating in different clusters/pods, where the ratio exceeds 1:1. The bandwidth contesting leads to switch buffers overloading, which then in turn start losing packets [23]. In addition, the buffer queuing and processing delay at the multi-tier switches bring large latency for the inter-cluster traffic that a packet needs to traverse the aggregation and core switches to reach its destination. Another challenge introduced with hierarchical network topology is the lack of network fault tolerance, especially at the core switching layer resulting from the lower physical connectivity. Hardware (switch) failures in aggregation of core switching layers will significantly deteriorate overall network performance.

To interconnect the multi-tier switching nodes, the electrical-to-optical-to-electrical (O/E/O) conversions exist between switching layers, which thereby significantly increases the number of transceivers and, hence, cost and power consumption [57]. Instead of the low-rate on-off keying (OOK) modulation, multi-level modulation like pulse-amplitude modulation (PAM)-4 schemes are gradually employed in the data centers fueled by the demand for higher data-rate [58]. To process these format-dependent signals, dedicated parallel optics and electronic circuits are required at the front-end of the electronic switches. This introduces extra cost and power consumption in the hierarchical network topology.

Table 1 – Comparison of optical switching technology.

| Switching Technology | OCS | OBS | OPS |
|---|---|---|---|
| Switching Granularity | coarse | medium | fine |
| Flow Completion Delay | high | medium | low |
| Bandwidth Utilization | low | medium | high |
| Control Overhead | low | low | high |
| Complexity | low | medium | high |
| Applicability | medium | low | high |

**Optical Data Center Networks**

Optical switching, as a future-proof solution to overcome the bandwidth bottleneck of electrical switches, has attracted the widespread attention to researchers. Due to the optical transparency, switching the data in the optical domain is independent of the bit-rate and data-format of the traffic. Thus, optical switching supports much higher bandwidth than electrical switching and at much lower packet completion time due to the removing of electronic circuits for switching. In addition, WDM technology can be employed to boost the optical network capacity at a superior power-per-unit bandwidth performance. Combining with the WDM, optical switching is a viable solution to overcome the count limitation of high-speeds pins and front panel connectors at electrical switches. Moreover, the optical switches do not require any power-consuming E/O and O/E conversions, which significantly reduces the number of expensive and power-hungry transceivers. All these benefits can be exploited to flatten the network topology and thus sidestepping the scaling wall of the hierarchical data center topology.

*2.1 Optical Switching Technologies*

To date, three main optical switching technologies have been investigated which resulted in increasing data transfer capabilities for the data center networks.

*Optical Circuit Switching (OCS):* OCS has three distinct steps: links set-up, data transmission and links tear-down. One of the main features of OCS is its two-way reservation process in the phase of link circuit set-up, where a source sends a request for setting up a circuit and then receives an acknowledgement back from the corresponding destination [59]. The overall transfers suffer from long set-up times relative to connection holding time, seriously deteriorating the network throughput. In addition, all data transmission of a connection in OCS network follow the same path, no statistical multiplexing of the client packet can be achieved at any intermediate node. More specifically, bandwidth allocation is a coarse granularity, which is allocated by one wavelength at a time. However, most modern applications in practice require the sub-wavelength connectivity and these high-bit-rate applications often involve "traffic bursts" that last only a few milliseconds or less. Furthermore, since no optical buffer existing, the capacity of the circuit link must equal the peak data rate, which can be orders of magnitudes higher than the average data rate, for bursty sources [60]. This results in the low bandwidth utilization for OCS-based data centers, especial for networks with many communication pairs with bursty traffic patterns.

*Optical Burst Switching (OBS):* In OBS-based data center schemes, the source nodes first send a burst header (control packet) on a separate control link (similar to the link set-up step of OCS) to reserve the optical bandwidth with the configuration of switches along an optical path for the burst forwarding of optical payload. The OBS scheme, unlike OCS, can send out the payload on a data channel without having to receive the response signal first. After sending out the burst payload, another control signal (similar to the link tear-down of OCS) is sent out to release the reserved optical bandwidth [61]. This implies that the offset time T between the burst header and the burst packets can be much less than the circuit set-up time, improving the network throughput. Benefitting from the one-pass bandwidth reservation for the duration of actual data transfer, the OBS paradigm provides the sub-wavelength switching granularity. However, due to OBS schemes generally do not need optical buffer, a big issue related to the one-way reservation OBS is how to deal with packet contention and prevent the contention caused packet dropping [62]. Another challenge of OBS related to the long-time bandwidth reservation and the using of a non-zero offset time is the high payload completion time encountered by each burst communication, not fully supporting the bursty traffic patterns as well [63].

*Optical Packet Switching (OPS):* OPS paradigm uses in-band control information where the header or label follows the rest of the packet payload closely, so there is no reservation possible (thus decreasing the end-to-end latency) and the bandwidth can be utilized in the most flexible way [64]. Due to OPS scheme allows statistical sharing of the optical bandwidth among packets belonging to different source and destination pairs, OPS scheme is thus suitable for supporting burst traffic scenarios. The packet payloads in OPS-based data centers remain in the optical domain, while the header or label may be electronically or optically processed (though the optical logic is very primitive). The generating and processing of fast header rely heavily on optical labeling techniques. To keep the percentage of the control overhead down, OPS-based data centers normally employ fast (nanoseconds reconfiguration time) optical switches based on semiconductor optical amplifiers (SOAs) or arrayed waveguide grating routers (AWGRs). Contention resolution is typically achieved by a combination of wavelength conversion, fiber-optic delay lines (FDLs) and, in rare cases, deflection routing [65]. Exploiting the WDM technique, the OPS paradigms significantly improve the network capacity where multiple streams of packets are multiplexed in the wavelength domain [66]. Benefitting from these features, OPS technologies offer a suitable solution for data center applications which requires on-demand transmitting the bursty and small data sets.

Table 1 summarizes the characteristics of these three optical switching technologies. The OCS has coarse switching granularity and thus resulting in high packet completion time and low bandwidth utilization, but it benefits the low implementation complexity that is easier to be deployed. As a comparison, the OPS, benefitting from the fine switching granularity, can fast complete the traffic flow with high bandwidth utilization. However, the implementation complexity and high control overhead limit its practical application in large-scale optical DCNs. As a compromise, the characteristics of OBS lies between OCS and OPS technologies.

### 2.2 Optical Switches: State-of-art

To date, exploiting various building blocks, many solutions employing optical switches have been investigated [67], such as 3D micro-electrical mechanical switches (MEMS), Liquid Crystal on Silicon (LCoS) display matrices, micro-ring resonators (MRRs), Mach-Zehnder interferometers (MZIs), tunable lasers and arrayed waveguide grating routers (AWGRs), and semiconductor optical amplifiers (SOAs). Determined by the exploited building blocks, switch reconfiguration time, as the main switch investigate parameter, can vary three orders of magnitude from milliseconds to nanoseconds, determining the granularity of a switch and therefore, its application.

*Slow (milliseconds and microseconds) Optical Switches:* Micro-electrical mechanical switches (MEMS) switches are micrometer-scale devices that rely on mechanical moving micro-mirrors to switch the optical signal from input ports to output ports [68]. An array of $N^2$ micro-mirrors is needed to build a N×N direct switching MEMS switch. The area to lay the mirrors out and the mirror size limits the port-radix to N≤32. By steering a pair of micro-mirrors, indirect switching MEMS is implemented in a three-dimensional (3D) plane [69], which requires only 2N mirrors configurable to N discrete angles. The switch reconfiguration time (milliseconds magnitude) is strongly determined by the mirror response speed to the precise movement between these N discrete angles. Large scale (N=1100) 3D-MEMS switches [70] have been investigated with a 4 dB maximum insertion loss and commercial product with 25 ms reconfiguration time supports up to N=320 with a 3 dB maximum insertion loss [71].

Liquid crystal (LC) technology can be used to switch light based on the LC's birefringence to control the polarization of incident light [72]. For the liquid crystal on silicon (LCoS) technique, a large size of 1×N optical switch can be built by depositing a reflective array of LC cells on a silicon backplane [73]. At the LCoS optical switch, all channels focus on a pixel area depositing multiple phase-shifting LCs that collectively form a controllable linear phase grating (in the range 0 to $2\pi$) to steer a reflected channel to an output port. Since each channel is switched separately, multiple channels can be selected and multiplexed into the same output port. Multiple 1×N switches can be stacked in a Spanke topology for N×N switching [74]. The reconfiguration time for this switch type is of the order of tens of milliseconds [75]. [76] demonstrates a 1×40 LCoS switch with 8 dB insertion loss and -40 dB crosstalk.

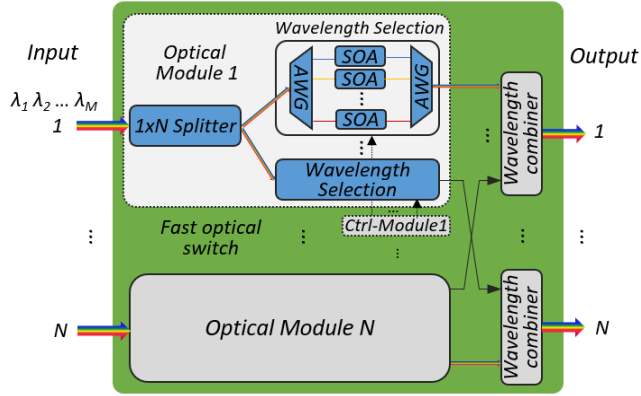

Figure. 4. Broadcast-and-select (B&S) topology.

Micro-ring resonator (MRR) is waveguide topology that can exploit the thermo-optic (to) effect for optical switching. A circular resonant cavity of micrometer-scale radius (r) can be formed by bending back a waveguide onto itself. When the optical path of the ring is an integer multiple (L) of the guided wavelength ($\lambda$), the ring starts resonating according to the principle of $(2\pi r)\cdot R_i = L\cdot\lambda$, where $R_i$ is the effective refractive index of the ring [77]. For switching, multi-rings are placed between two bus waveguides for an optical signal, at the resonant wavelength, to be coupled from one waveguide to the other. Heating a ring changes its effective refractive index and shifts its resonant wavelength, thereby enabling wavelength-selective switching [78]. A silicon photonic 8×8 switch, reconfigurable in the order of a few microseconds, is recently reported [79] with -16.75 dB average crosstalk and 8.4 dB average insertion loss.

*Fast (nanoseconds) Optical Switches:* A Mach-Zehnder interferometer (MZI) waveguide exploiting electro-optic (EO) effect can perform fast switching by changing the waveguide refractive index [80]. In a 2×2 MZI switch unit, a coupler divides the input optical signal to the two MZI arms. In the "on" state, the two arms are in anti-phase leading to constructive interference at the through waveguide. In the "off" state, the two MZI arms are in phase and a signal entering an input waveguide leads to destructive interference at the crossover waveguide, switching the input signal [81]. To date, the largest demonstrated silicon integrated MZI switch is 32×32 built in Benes topology with reconfiguration time of 1.2 ns and the high insertion loss (20.8 dB) and crosstalk (-14.1 dB) [82].

A passive router can be built based on a N×N arrayed waveguide grating (AWG), assuming wavelength tuning ($N^2$ fixed wavelength transceivers or N wavelength tunable transceivers) is used at the input/output ports [83]. In an AWG router (AWGR), each input port at the same time exploits the same N wavelengths to establish a strictly non-blocking all-to-all connectivity [84]. Following the cyclic mechanism, the wavelengths from two adjacent input ports appear at the output ports cyclically rotated by one position. Hence, each output port could receive N wavelength channels, one from each input port. The input waveguides are spaced such that, on any phased array waveguide, signals of the same wavelength from N input ports, have an additional phase difference. Signals are separated again at the output coupler and directed to different output ports. Switch reconfiguration time, which can be less than 10 ns, is determined by the time required for wavelength tuning [85]. A silicon AWGR with N=512 has been fabricated with a high inter-channel crosstalk of -4 dB [86].

Semiconductor optical amplifier (SOA) can be used as an optical switch gate, the "on" state providing broadband amplification and the "off" state blocking the incident signal [87]. Based on a kind of broadcast-and-select (B&S) topology

Table 2 – Comparison of optical switches.

| Switching technology | Switching time | Switch scale | Insertion loss | Crosstalk |
| --- | --- | --- | --- | --- |
| 2D MEMS | O(100)μs | 32×32 | low | low |
| 3D MEMS | O(10)ms | 1100×1100 | low | low |
| LC | O(100)μs | 2×2 | low | low |
| LCoS | O(10) ms | 1×40 | medium | low |
| TO-MRR | O(1)μs | 8×8 | low | medium |
| EO-MZI | O(1)ns | 32×32 | high | high |
| AWGR | O(1)ns | 512×512 | medium | high |
| SOA | O(100)ps | 16×16 | low | low |

Medium level: 5dB ⩽ Insertion Loss ⩽ 10dB and −40dB ⩽ Crosstalk ⩽ −20dB

as shown in Fig. 4, N×N SOA-based optical switches can be implemented, where each switching path exploits an SOA gate. The input signal is split into N paths, in the B&S architecture, for broadcasting the input signal to all output ports. N parallel SOA gates are used to switch ("on" or "off") the N paths signal, establishing an input-output port connection. The broadcast operation of B&S architecture allows wavelength, space and time switching, and another benefit is that the SOA gain in "on" state inherently compensates the splitting and combining losses [88]. Additionally, the high "on"/"off" extinction ratio brings its excellent crosstalk suppression. Nevertheless, the scalability in the B&S structure is limited by the splitter caused power loss and the high number of waveguide crossing [89]. Multi-stage topology can be used to arrange smaller B&S modules to scale out the switch to a large size, but the scalability is still limited by the amplified spontaneous emission (ASE) noise. Lossless, integrated 16×16 SOA switches with nanoseconds reconfiguration time have been fabricated [90, 91], based on smaller B&S switching modules interconnected in a three-stage Clos architecture. These features make SOA-based switch a suitable candidate for data center applications which requires fast and high-bandwidth transmission.

*2.3 Optical Data Center Network: State-of-art*

Various optically switched architecture prototypes, based on the above optical switches, have been proposed to demonstrate the potential of optical data center networks. Optical data center networks are mainly classified into two categories based on the switching techniques used, the electrical/optical hybrid scheme, where electrical along with the optical switches constitute together for the network interconnection, and the full optical scheme, where only optical switches (slow or fast) are employed.

*Hybrid Electrical/Optical Data Center Networks:* A number of hybrid electrical/optical interconnecting architectures have been reported for data centers, such as Helios [92], HydRA [93], c-Through [94] and RotorNet [95]. The electrical switches typically connect all the servers in a multi-level hierarchy and short reconfiguration time with large connectivity to handle small-size and bursty traffic patterns. The ToR are interconnected using both this down-link electrical packet-switched network and an up-link optical circuit-switched network. The optical circuit-switched network, implemented by a single or an array of slow optical switches like MEMS, provides large capacity links for high-volume and slow-changing traffic.

All the aforementioned prototypes need a centralized network scheduler to reconfigure the entire architecture, in response to the traffic dynamics. The central schedule requires network-wide demand estimation and resource schedule [95, 96]. Apart from the long optical switch (MEMS) reconfiguration time, ranging from microseconds to tens of milliseconds, the schedule and control introduce a significant extra processing latency. For instance, in order to achieve high resource utilization in Helios and c-Through, Edmond algorithm needs hundreds of milliseconds to converge to a network-wide matching [94]. Hence, the proposed architectures are better suited for non-bursty applications such as data migration and storage backup [93] where traffic is aggregated last more than a couple of seconds [92], to compensate the reconfiguration overhead.

*All Optical Data Center Networks:* This category includes the proposals exploiting the slow (sub-millisecond/milliseconds) or fast (nanoseconds) optical switches. OSA and Proteus [97, 98], as the all-optical switching architectures are proposed, where the ToRs are connected to a central MEMS switch, through optical Mux/Demux. A key challenge here is the slow reconfiguration time including both hardware switching time (microseconds to tens of milliseconds) and controlling overhead (hundreds of microseconds to seconds). Mordia [96], Wavecube [99], RODA [100] and OPMDC [101] are all built based on wavelength selective switches utilizing either a ring topology or a multi-dimensional cube structure. The limited port-count of wavelength selective switch requires stacking and cascading multiple switches, deteriorating network performance in terms of packet loss, flexibility and latency [102].

Fast network interconnecting like IRIS [103], DOS [104], Petabit [105], LIONS [106] and Hi-LION [107, 108] are reported based on AWGRs, of which the Petabit proposed a three-stage Clos network and Hi-LION demonstrates a mesh-like network exploiting both local and global AWGRs. The network performance and interconnecting scalability of the aforementioned proposals are strongly depended on the port radix [84] and the capability of wavelength tuning components such as tunable lasers or tunable wavelength converters. In addition, the wavelength-related operation as a block on the road to deploy WDM technology further limits the network capacity. An all-optical network, Baldur, is proposed in [109] based on transistor laser (TL) to enable high-speed and power-efficient communications in computing systems. However, the complex current control for the large radix TL array limits the practical deployment in the computing network. SOAs working as switching gates, the OPSquare [110], HiFOST [111], Vortex [112], ROTOS [113] and OSMOSIS [114] are proposed with SOA-based B&S, featuring of nanoseconds switching time. OPSquare uses a parallel-module switch architecture with distributed control and scales by adding more modules and wavelengths. There is only one switching stage, irrespective of the port-count, implemented in the modules. At most two-hop is enough for the traffic forwarding between any two different edge nodes [115]. Utilizing WDM and B&S stages, the OSMOSIS equips high-capacity and low-latency forwarding of the synchronously arrived fixed-length optical packets. Scaling is limited by the data plane and control plane complexity [114]. The scheduler implementation spans multiple interconnected chips, increasing not only network complexity but also the latency.

Given the bursty traffic features and high fan-in/out hotspots patterns in data center networks, slow optical switches providing static-like and pairwise interconnections would only be beneficial as supplementary switching elements. In contrast, fast optical switches with nanosecond-reconfigurable time can handle arbitrary traffic and can be deployed at any layer of the data center network. Considering this, fast optical switches-based network topologies supporting nanoseconds optical packet switching offers a potentially future-proof solution for the fast and high-capacity data center networks.

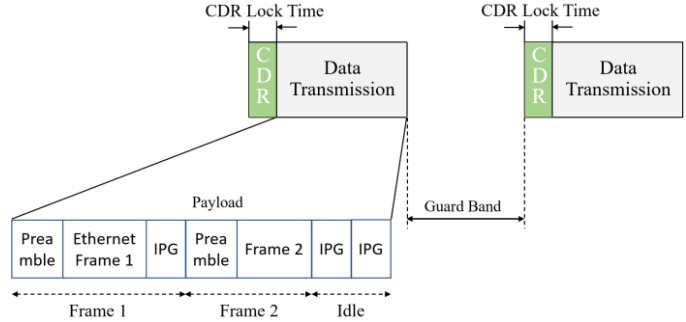

Figure. 5. Clock and data recovery step at the receiver.

*2.4 Technical Challenges*

The network and optics communities have been extensively investigating the fast (nanoseconds) optical switching techniques for many years. However, each community tries to address problems and challenges from their own perspective. For instance, the optics communities focus on developing technologies for single components and devices that achieve nanosecond switching configuration time while devoting little attention to solving the network interconnecting challenges, e.g., flexible quality of service (QoS) provisioning or scalable scheduling. In contrast, the networking community proposed a number of solutions that can scale the network interconnecting and address the system challenges such as demand estimation or network bandwidth reconfiguration [116]. Thus, the proposed solutions need to be integrated and, in some cases, need to be redesigned for the cases proposed independently by each community. Following, the main technical challenges that limit the practical deployment of optical data center networks are summarized.

*Fast and Scalable Switch Control:* Despite the promises held by fast optical switches, the corresponding nanoseconds-scale control mechanisms are required to control the switches and to fast forward the data traffic [117]. Typically, an optical label or header, carrying the destination information of the data packet, is associated with the optical data packet to be processed at the specific switch controller. Based on the received label matrix, the switch controller computes the optical-switch configuration and then accordingly forward the data packets [118]. Short switch configuration time, consisting of both controlling overhead and hardware switching time, is essential as it determines the network latency and throughput performance [119]. Thus, to fully utilize the nanoseconds-magnitude hardware switching time and to reduce the flow completion time of short data packets, the switch controller needs to process the label signals and configure the switch within nanoseconds [120]. Moreover, the switch controlling overhead should be independent of the network scale. Considering the scale of practical DCNs, which typically comprise hundreds of thousands servers, switch control mechanism needs to be performed in parallel and independently for every optical switch, not a network-wide scale schedule [121]. In addition, for the label control mechanism, the edge nodes should be time-synchronized connected to the optical switches at a very fine granularity (ideally few nanoseconds) to align the label signals and corresponding data packets [4]. Even any time inaccuracy in the synchronization phase can be accordingly compensated with a customized interpacket gap, however, this would reduce the overall network throughput. Therefore, the implementation of fast and scalable switch control requires a nontrivial amount of ingenuity and custom hardware support.

*Lack of Optical Memory:* The lack of optical buffer is one of the main fundamental differences between optical switch and electrical switch. Electrical switches typically employ random access memories (RAM) to buffer the packets that lost contention. Due to the lack of effective RAM in the optical domain, the conflicted packets at the optical switch would be dropped, thereby resulting in packet loss [122]. Thus, packet contention resolution is another unsolved challenge that needs to be solved to guarantee fast switch control. Several solutions have been proposed to address such issue, based either on wavelength conversion [123], optical fiber delay lines (FDLs) [124] or deflection routing [125]. However, none of them is practical for large-scale DCNs, due to the extra hardware deployment of wavelength conversion, fixed buffering time of FDLs and the management complexity of deflection routing. Label control mechanism enabling dropped packet retransmission provides a promising solution to address the packet contention caused packet loss, combining the deployment of RAM at the edge nodes (such as ToRs) [126]. To minimize the introduced retransmission delay, the optical switch should be employed relatively close to the edge nodes, which requires to flat the network topology. Moreover, an efficient scheduler is essential to intrinsically reduce the time-consuming packet retransmission.

*Fast Clock Data Recovery (CDR):* Unlike the synchronized point-to-point connections between any paired ports in an electrical switch, the optical switch creates momentary physical links between source and destination ports [127]. Therefore, in an optical packet switching network, where the clock frequency and phase of data packets vary packet by packet, new physical connections are created every time reconfiguring the optical switch [128]. The receivers thus have to continuously adjust the local clock (consisting of both frequency and phase) to properly sample the incoming optical packets and thereby recovering the payload, as illustrated in Fig. 5. As no payload data can be valuably received before the CDR completing, the long CDR processing time (hundreds of nanoseconds for off-the-shelf transceivers) will deteriorate the network throughput, especially in the intra data center scenarios where applications and services produce short traffic packets [129]. In [9], the clock phase caching technique is used to accelerate the CDR processing at the receiver side. However, to adjust the clock phase interpolator for every packet, extra time at the transmitter side is required which significantly increases the interpacket gap time and whereby offsets the efficacy of short CDR time. Burst-mode receivers enabling nanoseconds CDR processing time based on over-sampling or gated oscillators have been extensively investigated in passive optical networks [130]. These burst-mode techniques, however, introduce the design complexity and increase the deployment cost. These burst-mode receivers also need to be re-evaluated aiming for higher (>25 Gb/s) data rates, not suitable for DCNs with 100Gb/s links to be deployed.

*Reconfigurable Connectivity:* The rapid emerging of multi-tenant applications and services with mix data flows impose their own set of various requirements, such as bandwidth and forwarding priority, to the network infrastructure. These specific requirements dynamically change as the application switch-over. Thus, the reconfigurable and highly flexible connectivity is required in order to optimize the hardware resources, overcoming the limitation of currently deployed static and semi-automated control frameworks [116, 131]. One promising strategy is to virtualize the infrastructures in a fully operable and flexible way by the software-defined networking (SDN) to enable such reconfigurable environments [115, 132]. The SDN control plane exploits an open, standard, vendor-independent and technology-transparent southbound interface to monitor and configure the underlying data plane. To facilitate the integration of SDN control plane, the components of data plane such as ToRs and optical switches should be compatible with the open interface exhibited by the SDN control plane. Given the specificity and characteristic of the optical switching network, proper extensions and customization on the open protocols (e.g., OpenFlow [133]) have to be performed and implemented. Moreover, efforts need to be made to develop various functional engines running at the SDN control plane to flexibly virtualize the optical infrastructure, supplying reconfigurable connectivity.

## Conclusion

Variety of applications and services depend on data centers to provide the high reliability and availability of computing and storage resource at minimal costs. Along with the emerging of traffic boosting applications, the bandwidth bottleneck of electrical switches forces the migration of switching from the electrical domain to the optical domain. Data centers are thus moving towards full optical switching with technical evolutions of both optical switches and network topologies to satisfy the demands of massively increasing data center traffic. In this paper, we have presented a summary of data centers traffic characteristics and topologies trends in data center networks which are based on electrical switches. The shortcomings and challenges for electrically switched data centers are also reported to reveal the trends of full optical switching. To that end, we present a brief summary of optical switching technologies that will enable ultra-high bandwidth links, in addition to an overview of optical network topologies that will enable the high utilization of bandwidth and thereby lower cost and power consumption. The full optical switching is expected to deploy in data centers in the next decade, enabling the developments of new applications like artificial intelligence and machine learning, as well as providing the fast, reliable and cost-effective services to users.

## Funding

This work is supported by Natural Science Foundation of China (62101065, 62171059, 62125103, 62171175). Fund of State Key Laboratory of Information Photonics and Optical Communications (Beijing University of Posts and Telecommunications) (No. IPOC2021ZT08), P. R. China. National Key R&D Program of China (2018YFB1801702).